\documentclass[12pt]{article}
\usepackage{amsmath}
\usepackage{color}
\usepackage{graphicx}
\begin{document}
\baselineskip=18 pt
\begin{center}
{\large{\bf Generalized KG-oscillator with a uniform magnetic field under the influence of Coulomb-type potentials in cosmic string space-time and Aharonov-Bohm effect }}
\end{center}

\vspace{0.5cm}
\begin{center}
{\bf Faizuddin Ahmed}\footnote{\bf faizuddinahmed15@gmail.com ; faiz4U.enter@rediffmail.com}\\
{\bf National Academy Gauripur, Assam, 783331, India}
\end{center}
\vspace{0.5cm}

\begin{abstract}

We solve a generalized Klein-Gordon oscillator (KGO) in the presence of a uniform magnetic field including quantum flux under the effects of a scalar and vector potentials of Coulomb-types in the static cosmic string space-time. We obtain the energy and corresponding eigenfunctions, and analyze a relativistic analogue of the Aharonov-Bohm effect for bound states.

\end{abstract}


\vspace{0.3cm}

{\bf PACS Number(s):} 03.65.Pm, 03.65.Ge, 98.08.Cq

\section{Introduction}

The relativistic quantum dynamics of spin-$0$ massive charged particles with electromagnetic field have been of current interest in theoretical physics. The electromagnetic interactions are introduced into the KG-equation through the minimal substitution, $p_{\mu}\rightarrow p_{\mu}-e\,A_{\mu}$ where, $e$ is the electric charge and $A_{\mu}$ is the four-vector potential of the electromagnetic field \cite{LDL,VB,WG,DJG}. Several authors have studied with the electromagnetic interactions in the relativistic quantum systems \cite{ZW,R1,R2,EPJP,ERFM}.

The relativistic quantum dynamics of spin-$0$ in the background of curved space-time have been investigated by several authors, for example, presence of a magnetic field under the effects of Cornell-type potential in cosmic string space-time \cite{ERFM}, KG-oscillator in the presence of a magnetic field in cosmic string space-time \cite{AB}, KG-oscillator in cosmic string space-time \cite{HH}, DKP oscillator in cosmic string space-time \cite{HH3}, spin-zero system of DKP-equation in cosmic string space-time \cite{MdM}, DKP-equation for spin-zero bosons with a linear interaction in cosmic string space-time \cite{HH7}, effects of torsion and potential on KG-oscillator \cite{AHEP} and quantum dynamics of spin-$0$ particle \cite{AHEP4}, respectively in cosmic string space-time, non-inertial and torsion effects on KG-oscillator subject to potential in cosmic string space-time \cite{CJP}, KG-oscillator on curved background using the Kaluza-Klein theory \cite{JC}, and non-inertial effects on generalized DKP-oscillator in cosmic string space-time \cite{SZ}. So far, generalized KG-oscillator in the background of cosmic string space-time with an external field under the effects of potential hasn't yet investigated. In this work, we have chosen arbitrary function for the studies of generalized KG-oscillator with a magnetic field including quantum flux under the effects of scalar and vector potentials in cosmic string space-time, and observe an analogue of the Aharonov-Bohm effect.  

The Aharonov-Bohm effect \cite{MP,VBB} is a quantum mechanical phenomena that arises due to the presence of quantum flux $\Phi_B$ \cite{YA} in the topological defects space-times. This effect has investigated by many authors, such as, on position-dependent mass system with torsion effects in \cite{IJMPD}, in the context of Kaluza-Klein theory on curved background ({\it e. g.}, \cite{JC,AHEP2} and related references their in), and in the relativistic quantum mechanics ({\it e. g. }, \cite{R2,EPL2} and related references their in).

To obtain bound states solution of the wave-equations, various potentials have been used. The Coulomb-type potential which is our interest here has given much attention due to short range interaction. This kind of potential studies, in the propagation of gravitational waves \cite{HA}, in quark models \cite{CLC}, and in the relativistic quantum mechanics by several authors \cite{R1,R2,AHEP2,MH,EVBL,KB,LCNS,HH2}. This type of potential is given by
\begin{equation}
S (r)=\frac{k_c}{r}\quad,\quad V (r) =\frac{\xi_c}{r}=A_0
\label{1}
\end{equation}
where $k_c,\xi_c$ are the Coulombic confining parameters. Here $S(r)$ is the scalar potential and $V (r)$ is the fourth component of the electromagnetic vector potential. We have restricted here $|S (r)| > |e\,V (r)|$.

In Ref.\cite{AB}, KG-oscillator with and/or without a magnetic field in the cosmic string space-time was investigated. In this work, we introduce a scalar potential $S(r)$ by modifying the mass term $M \rightarrow (M+S)$ in the equation, and the fourth component $A_0$ of electromagnetic vector potential in the equation. We solve generalized KG-oscillator by choosing different function $f (r)$ in the presence of a magnetic field including quantum flux under the influence of Coulomb-type potentials in cosmic string space-time, and analyze a relativistic analogue of the Aharonov-Bohm effect for bound states.

\section{Generalized KGO under the effects of Coulomb-type potentials in cosmic string space-time and AB-effect }

The static cosmic string space-time in polar coordinates $(t, r, \phi, z)$ is described by the following line element \cite{ERFM,AB,HH,HH3,MdM,HH7,AV} :
\begin{equation}
ds^2=-dt^2+\alpha^2\,r^2\,d\phi^2+dr^2+dz^2,
\label{2}
\end{equation}
where $\alpha=1-4\,\mu$ is the wedge parameter, and $\mu$ is the linear mass density of the string. In the cylindrical symmetry, we have that $ 0 < r < \infty$, $ 0 \leq \phi <  2\,\pi$ and $-\infty < z < \infty$. In gravitation and cosmology, the parameter $\alpha$ assumes values in the interval $ 0 < \alpha < 1$ \cite{MOK,CF}. The line element (\ref{2}) describes the geometry with a conical singularity which corresponds to the curvature tensor $R^{r,\phi}_{r,\phi}=(\frac{1-\alpha}{4\,\alpha})\,\delta^2 (\vec{r})$, with $\delta^2 (\vec{r})$ the two-dimensional delta function. This conical singularity is such that the curvature is concentrated on the axis, and in all other points, the curvature vanishes \cite{DDS}.

The relativistic quantum dynamics of spin-$0$ massive charged particle with a scalar potential $S(r)$ is described by the following KG-equation \cite{ZW,R1,R2,ERFM,CJP,IJMPD,AHEP,AHEP4,AHEP2}
\begin{equation}
\left [\frac{1}{\sqrt{-g}}\,D_{\mu}(\sqrt{-g}\,g^{\mu\nu}\,D_{\nu})-(M+ S(r))^2 \right]\,\Psi=0.
\label{3}
\end{equation}
where the minimal substitution is defined by $D_{\mu}=\partial_{\mu}-i\,e\,A_{\mu}$, $M$ is the rest mass of the particle.

We consider the four-vector potential $A_{\mu}=(-V,\vec{A})$ with three vector-potential in symmetric gauge is defined by
\begin{equation}
\vec{A}=(0,A_{\phi},0).
\label{5}
\end{equation}
Now, let us assume that the topological defects have an internal magnetic flux field (with magnetic quantum flux $\Phi_B$) \cite{YA,MSC,CJP}. The three-vector potential in symmetric gauge is given by \cite{R2,AHEP,AHEP4,IJMPD,AHEP2,EPL2}:
\begin{equation}
A_{\phi}=-\frac{1}{2}\,\alpha\,B_0\,r^2+\frac{\Phi_B}{2\,\pi}.
\label{6}
\end{equation}
Here the applied magnetic field $\vec{B}=\nabla \times \vec{A}=-B_0\,\hat{k}$ is along the $z$-axis. 

For the geometry (\ref{2}), the KG-equation (\ref{3}) becomes
\begin{eqnarray}
&&[-\left(\frac{\partial}{\partial t}+i\,e\,V \right)^2+\frac{1}{r}\,\frac{\partial}{\partial r}\,(r\,\frac{\partial}{\partial r})+\frac{1}{\alpha^2\,r^2}\,\left(\frac{\partial}{\partial \phi}-i\,e\,A_{\phi} \right)^2+\frac{\partial^2}{\partial z^2}\nonumber\\
&&-(M+S)^2]\,\Psi=0.
\label{7}
\end{eqnarray}

To couple generalized oscillator with KG-field, following change in the radial momentum operator is considered \cite{AHEP2,EPL,SZ}:
\begin{equation}
\vec{p} \rightarrow \vec{p}-i\,M\,\Omega\,f(r)\,\hat{r} \quad \mbox{or}\quad \partial_{r} \rightarrow \partial_{r}+M\,\Omega\,f(r),
\label{osc}
\end{equation}
where $\Omega$ is the oscillator frequency of the particle and $f(r)$ being an arbitrary function.

Therefore, equation (\ref{7}) becomes
\begin{eqnarray}
&&[-\left(\frac{\partial}{\partial t}+i\, e\, V \right)^2+\frac{1}{r}\left (\frac{\partial}{\partial r} + M\,\Omega\, f(r) \right)\,\left (r\,\frac{\partial}{\partial r}-M \,\Omega\,r\,f(r) \right)\nonumber\\
&&+\frac{1}{\alpha^2 r^2}\left(\frac{\partial}{\partial \phi}-i\, e\, A_{\phi} \right)^2+\frac{\partial^2}{\partial z^2}-(M+S)^2 ]\,\Psi (t,r,\phi,z)=0.
\label{osc2}
\end{eqnarray}

Since the metric is independent of the coordinates $(t, \phi, z)$, it is reasonable to write the solution to Eq. (\ref{osc2}) as
\begin{equation}
\Psi (t, r, \phi, z)=e^{i\,(-E\,t+l\,\phi+k\,z)}\,\psi (r),
\label{8}
\end{equation}
where $E$ is the energy of the field, $l=0,\pm 1,\pm 2,....$ are the eigenvalues of the $z$-component of the angular momentum operator, and $k$ is a constant.

Substituting the solution (\ref{8}) into the Eq. (\ref{osc2}), we obtain the following radial wave-equation for $\psi (r)$ :
\begin{eqnarray}
&&\psi''(r)+\frac{1}{r}\psi' (r)+[(E-e V)^2-(M+S)^2-k^2-\frac{(l-e A_{\phi})^2}{\alpha^2 r^2}\nonumber\\
&&-M^2\,\Omega^2\,f^2-M\,\Omega\,\left(f'+\frac{f}{r}\right)]\psi (r)=0.
\label{9}
\end{eqnarray}

In this work, we have chosen few arbitrary function for the studies of generalized KG-oscillator as follows:

\subsection{Cornell-type function : $f (r)= b_1\,r+\frac{b_2}{r}$}

Here we have chosen a Cornell-type potential form function where, $b_1, b_2$ are arbitrary constants.

Substituting the potentials (\ref{1}) and (\ref{6}) into the Eq. (\ref{9}) using the above Cornell-type function, we obtain :
\begin{equation}
\left [\frac{d^2}{dr^2}+\frac{1}{r}\,\frac{d}{dr}+\lambda-M^2\,\omega^2\,r^2-\frac{j^2}{r^2}-\frac{a}{r} \right]\,\psi (r)=0,
\label{10}
\end{equation}
where
\begin{eqnarray}
&&\lambda=E^2-M^2-k^2-2\,M\,\omega_c\,l_0-2\,M\,\Omega\,b_1-2\,M^2\,\Omega^2\,b_1\,b_2,\nonumber\\
&&\omega=\sqrt{\Omega^2\,b_1^{2}+\omega^2_{c}},\nonumber\\
&&j=\sqrt{l^2_{0}+M^2\,\Omega^2\,b^2_{2}+k^2_{c}-e^2\,\xi^2_{c}},\nonumber\\
&&l_0=\frac{l-\Phi}{\alpha},\nonumber\\
&&\Phi=\frac{\Phi_B}{(2\pi/e)},\nonumber\\
&&a=2\,(e\,\xi_c\,E+M\,k_c)\nonumber\\\mbox{and}
&&\omega_c=\frac{e\,B_0}{2\,M}
\label{11}
\end{eqnarray}
is called the cyclotron frequency of the moving particle in the magnetic field.

Transforming $x=\sqrt{M\,\omega}\,r$ into the Eq. (\ref{10}), we obtain the following wave equation:
\begin{equation}
\left [\frac{d^2}{dx^2}+\frac{1}{x}\,\frac{d}{dx}+\zeta-x^2-\frac{j^2}{x^2}-\frac{\eta}{x} \right]\,\psi (x)=0,
\label{13}
\end{equation}
where
\begin{equation}
\zeta=\frac{\lambda}{M\,\omega}\quad,\quad \eta=\frac{a}{\sqrt{M\,\omega}}.
\label{14}
\end{equation}

Suppose the possible solution to Eq. (\ref{13}) is
\begin{equation}
\psi (x)=x^{j}\,e^{-\frac{x^2}{2}}\,H (x),
\label{15}
\end{equation}
where $H (x)$ is an unknown function. 

Substituting the solution (\ref{15}) into the Eq. (\ref{13}), we obtain
\begin{equation}
H'' (x)+\left [\frac{1+2\,j}{x}-2\,x \right]\,H' (x)+\left [-\frac{\eta}{x}+\zeta-2-2\,j \right]\,H (x)=0.
\label{16}
\end{equation}
Equation (\ref{16}) is the biconfluent Heun's differential equation \cite{ZW,R1,R2,IJMPD,AHEP2,ERFM,AR,SYS} with $H(x)$ is the Heun polynomials function.

The above equation (\ref{16}) can be solved by the Frobenius method. Writing the solution as a power series expansion around the origin \cite{GBA}:
\begin{equation}
H (x)=\sum^{\infty}_{i=0}\,c_{i}\,x^{i}.
\label{18}
\end{equation}
Substituting the power series solution (\ref{18}) into the Eq. (\ref{16}), we get the following recurrence relation for the coefficients:
\begin{equation}
c_{n+2}=\frac{1}{(n+2)(n+2+2\,j)}\,[\eta\,c_{n+1}-(\zeta-2-2\,j-2\,n)\,c_{n}].
\label{19}
\end{equation}
And the various coefficients are
\begin{eqnarray}
&&c_1=\left (\frac{\eta}{1+2\,j} \right)\,c_0,\nonumber\\
&&c_2=\frac{1}{4\,(1+j)}\,\left [\eta\,c_1-(\zeta-2-2\,j)\,c_0 \right].
\label{20}
\end{eqnarray}

The above power series expansion $H (x)$ becomes a polynomial of degree $n$ by imposing the following two conditions \cite{ZW,R1,R2,IJMPD,AHEP2,ERFM}
\begin{eqnarray}
\zeta-2-2\,j&=&2\,n,\quad (n=1,2,...)\nonumber\\
c_{n+1}&=&0.
\label{21}
\end{eqnarray}

By analyzing the first condition $\zeta-2-2\,j=2\,n$, we get the following expression of the energy eigenvalues $E_{n,l}$ :
\begin{eqnarray}
E^2_{n,l}&=&k^2+M^2+2\,M\,\omega\,\left(n+1+\sqrt{\frac{(l-\Phi)^2}{\alpha^2}+M^2\,\Omega^2\,b^2_{2}+k^2_{c}-e^2\,\xi^2_{c}}\right)\nonumber\\
&&+2\,M\,\omega_c\,\left(\frac{l-\Phi}{\alpha} \right)+2\,M\,\Omega\,b_1+2\,M^2\,\Omega^2\,b_1\,b_2.
\label{22}
\end{eqnarray}

The radial wave-function is given by
\begin{equation}
\psi_{n,l} (x)=x^{\sqrt{\frac{(l-\Phi)^2}{\alpha^2}+M^2\,\Omega^2\,b_2^{2}+k^2_{c}-e^2\,\xi^2_{c}}}\,e^{-\frac{x^2}{2}}\,H (x),
\label{function}
\end{equation}
where $H (x)$ is now a polynomial of degree $n$.

Note that Eq. (\ref{22}) does not represent the general expression of energy eigenvalue associated with $n^{th}$ radial mode for the bound states solution of the system. To find the individual energy level and wave-function for each radial mode, one must impose the recurrence condition, namely, $c_{n+1}=0$ on the eigenvalue problem as done in Refs. \cite{ZW,R1,R2,ERFM,IJMPD,AHEP2,AVV,JM}. For example, $n=1$, we have $\zeta-2-2\,j=2$ and $c_2=0$, which implies from Eq. (\ref{20})
\begin{eqnarray}
&&\eta\,c_1-2\,c_0=0\Rightarrow c_1=\frac{2}{\eta}\,c_0\Rightarrow \frac{\eta}{1+2\,j}=\frac{2}{\eta}\nonumber\\
&&\Rightarrow \omega_{1,l}=\left[\frac{a^2_{1,l}}{2\,M\,(1+2\,j)}\right]
\label{23}
\end{eqnarray}
a constraint on the parameter $\omega_{1,l}$. Note that its values changes for each quantum number $n$ and $l$, so we have labeled $\omega \rightarrow \omega_{n,l}$ and $a \rightarrow a_{n,l}$ since $a$ is related with $E_{n,l}$ in Eq. (11). Besides, we have adjusted the magnetic field $B^{1,l}_0$ so that Eq. (\ref{23}) can be satisfied and we have
\begin{equation}
\omega^{1,l}_c=\sqrt{\omega^2_{1,l}-\Omega^2}\leftrightarrow B^{1,l}_0=\frac{2\,M}{e}\,\sqrt{\omega^2_{1,l}-\Omega^2}.
\label{osc3}
\end{equation}

Thus, the radial mode $n=1$, we have from Eq. (\ref{22}) 
\begin{eqnarray}
&&E^2_{1,l}=k^2+M^2+2\,M\,\omega_{1,l}\,\left(2+\sqrt{\frac{(l-\Phi)^2}{\alpha^2}+M^2\,\Omega^2\,b^2_{2}+k^2_{c}-e^2\,\xi^2_{c}}\right)\nonumber\\
&&+2\,M\,\omega^{1,l}_c\,(\frac{l-\Phi}{\alpha})+2\,M\,\Omega\,b_1+2\,M^2\,\Omega^2\,b_1\,b_2.
\label{24}
\end{eqnarray}
Then, by substituting $\omega_{1,l}$ from Eq. (\ref{23}) and $\omega^{1,l}_c$ from Eq. (\ref{osc3}) into the Eq. (\ref{24}), it is possible to obtain the allowed value of the relativistic energy level $E_{1,l}$ of the lowest state. We can see that the lowest energy state given by Eqs. (\ref{23})--(\ref{osc3}) plus the expression in Eq. (\ref{24}) is defined for the radial mode $n=1$, instead of $n=0$. This effect arise due to the presence of Coulomb-type potentials in the quantum system. Also the wedge parameter $\alpha$ are in the ranges $0 < \alpha <1$, thus, the degeneracy of the relativistic energy eigenvalue is broken and shifted the energy level of the bound states.

The ground state total wave-function is given by
\begin{equation}
\Psi_{1,l}=x^{\sqrt{\frac{(l-\Phi)^2}{\alpha^2}+M^2\,\Omega^2\,b^2_{2}+k^2_{c}-e^2\,\xi^2_{c}}}\,{\sf e}^{\left[-\frac{x^2}{2}+i\,\left(-E_{1,l}\,t+l\,\phi+k\,z \right)\right]}\,\left(c_0+c_1\,x\right),
\label{25}
\end{equation}
where
\begin{eqnarray}
c_1&=&\left(\frac{\eta}{1+2\,j} \right)\,c_0\nonumber\\
&=&\sqrt{\frac{1}{\left(\frac{1}{2}+\sqrt{\frac{(l-\Phi)^2}{\alpha^2}+M^2\,\Omega^2\,b^2_{2}+k^2_{c}-e^2\,\xi^2_{c}}\right)}}\,c_0.
\label{26}
\end{eqnarray}

\subsection{Linear function : $f (r)=b_1\,r$} 

For a linear function, we have chosen $b_2 \rightarrow 0$ into the function $f(r)$ considered earlier. In that case, the radial wave-equation Eq. (\ref{9}) becomes
\begin{equation}
\left [\frac{d^2}{dr^2}+\frac{1}{r}\,\frac{d}{dr}+\lambda_0-M^2\,\omega^2\,r^2-\frac{j^2_{0}}{r^2}-\frac{a}{r} \right]\,\psi (r)=0,
\label{b1}
\end{equation}
where $a, \omega$ are given earlier and 
\begin{equation}
\lambda_0=E^2-M^2-k^2-2\,M\,\omega_c\,l_0-2\,M\,\Omega\,b_1\quad ,\quad j_0=\sqrt{l^2_{0}+k^2_{c}-e^2\,\xi^2_{c}}.
\label{b2}
\end{equation}

Following the similar technique as done earlier, one can obtain the energy eigenvalues $E_{n,l}$ associated with $n^{th}$ radial mode as follows :
\begin{eqnarray}
E^2_{n,l}&=&k^2+M^2+2\,M\,\omega\,\left(n+1+\sqrt{\frac{(l-\Phi)^2}{\alpha^2}+k^2_{c}-e^2\,\xi^2_{c}}\right)\nonumber\\
&&+2\,M\,\omega_c\,\left(\frac{l-\Phi}{\alpha} \right)+2\,M\,\Omega\,b_1,
\label{b3}
\end{eqnarray}
where $\omega=\sqrt{\Omega^2\,b^2_{1}+\omega^2_{c}}$.

The radial wave-function is given by
\begin{equation}
\psi_{n,l} (x)=x^{\sqrt{\frac{(l-\Phi)^2}{\alpha^2}+k^2_{c}-e^2\,\xi^2_{c}}}\,e^{-\frac{x^2}{2}}\,H (x),
\label{b4}
\end{equation}
where $H (x)$ is a polynomial of degree $n$.

As done earlier, one can find here also the individual energy level and wave-function for each radial mode. For the radial mode $n=1$, one will evaluate the energy level $E_{1,l}$ from the following equation
\begin{eqnarray}
&&E^2_{1,l}=k^2+M^2+2\,M\,\omega_{1,l}\,\left(2+\sqrt{\frac{(l-\Phi)^2}{\alpha^2}+k^2_{c}-e^2\,\xi^2_{c}}\right)\nonumber\\
&&+2\,M\,\omega^{1,l}_c\,(\frac{l-\Phi}{\alpha})+2\,M\,\Omega\,b_1.
\label{b5}
\end{eqnarray}
Since the wedge parameter $\alpha$ are in the ranges $0 < \alpha <1$, thus, the degeneracy of the relativistic energy eigenvalue is broken and shifted the energy level of the bound states.

The ground state total wave-function is given by
\begin{equation}
\Psi_{1,l}=x^{\sqrt{\frac{(l-\Phi)^2}{\alpha^2}+k^2_{c}-e^2\,\xi^2_{c}}}\,e^{\left[-\frac{x^2}{2}+i\,\left(-E_{1,l}\,t+l\,\phi+k\,z \right)\right]}\,\left(c_0+c_1\,x\right),
\label{b6}
\end{equation}
where
\begin{eqnarray}
c_1&=&\left(\frac{\eta}{1+2\,j} \right)\,c_0\nonumber\\
&=&\frac{1}{\sqrt{\left(\frac{1}{2}+\sqrt{\frac{(l-\Phi)^2}{\alpha^2}+k^2_{c}-e^2\,\xi^2_{c}}\right)}}\,c_0.
\label{b7}
\end{eqnarray}
With the following constraints
\begin{eqnarray}
&&\omega_{1,l}=\left[\frac{a^2_{1,l}}{2\,M\,(1+2\,j_0)}\right],\nonumber\\
&&\omega^{1,l}_c=\sqrt{\omega^2_{1,l}-\Omega^2}\leftrightarrow B^{1,l}_0=\frac{2\,M}{e}\,\sqrt{\omega^2_{1,l}-\Omega^2}.
\label{b8}
\end{eqnarray}

In Ref. \cite{AB}, authors studied the Klein-Gordon oscillator in the presence of an external uniform field in a cosmic string space-time without any kind of potential. The energy eigenvalue is given by
\begin{equation}
E^{2}_{n,l}=2\,m\,\sqrt{\Omega^2+\omega^2_{c}}\,(2\,n+\frac{l}{\alpha}+1)-2\,m\,(\Omega+\omega_c\,\frac{l}{\alpha})+m^2+k^2,
\label{b9}
\end{equation}
where $\Omega$ is the oscillator frequency.

Thus by comparing the energy eigenvalues (\ref{b3}) with the result (\ref{b9}), we can see that the result in this work get modified due to the presence of Coulomb-types scalar and vector potentials including the magnetic quantum flux which causes shifts the energy level and gives rise to an analogue of the Aharonov-Bohm effect for bound states.

\subsection{Coulomb-type function : $f(r)=\frac{b_2}{r}$}

We have chosen a Coulomb-type function by setting $b_1 \rightarrow 0$ into the function $f(r)$ considered earlier. In that case, the radial wave-equation Eq. (\ref{9}) becomes
\begin{equation}
\left [\frac{d^2}{dr^2}+\frac{1}{r}\,\frac{d}{dr}+\tilde{\lambda}-M^2\,\omega^2_{c}\,r^2-\frac{j^2}{r^2}-\frac{a}{r} \right]\,\psi (r)=0,
\label{c1}
\end{equation}
where $a, j$ are given earlier and
\begin{equation}
\tilde{\lambda}=E^2-M^2-k^2-2\,M\,\omega_c\,l_0.
\label{c2}
\end{equation}

Transforming $x=\sqrt{M\,\omega_c}\,r$ into the Eq. (\ref{c1}), we obtain the following wave equation:
\begin{equation}
\left [\frac{d^2}{dx^2}+\frac{1}{x}\,\frac{d}{dx}+\tilde{\zeta}-x^2-\frac{j^2}{x^2}-\frac{\tilde{\eta}}{x} \right]\,\psi (x)=0,
\label{c3}
\end{equation}
where
\begin{equation}
\tilde{\zeta}=\frac{\lambda}{M\,\omega_c}\quad,\quad \tilde{\eta}=\frac{a}{\sqrt{M\,\omega_c}}.
\label{c4}
\end{equation}

Substituting the solution (\ref{15}) into the Eq. (\ref{c3}), we obtain
\begin{equation}
H'' (x)+\left [\frac{1+2\,j}{x}-2\,x \right]\,H' (x)+\left [-\frac{\tilde{\eta}}{x}+\tilde{\zeta}-2-2\,j \right]\,H (x)=0.
\label{c5}
\end{equation}

Substituting the power series solution (\ref{18}) into the Eq. (\ref{c5}), we get the following recurrence relation for the coefficients:
\begin{equation}
c_{n+2}=\frac{1}{(n+2)(n+2+2\,j)}\,[\tilde{\eta}\,c_{n+1}-(\tilde{\zeta}-2-2\,j-2\,n)\,c_{n}].
\label{c6}
\end{equation}
And the various coefficients are
\begin{eqnarray}
&&c_1=\left (\frac{\tilde{\eta}}{1+2\,j} \right)\,c_0,\nonumber\\
&&c_2=\frac{1}{4\,(1+j)}\,\left [\tilde{\eta}\,c_1-(\tilde{\zeta}-2-2\,j)\,c_0 \right].
\label{c7}
\end{eqnarray}

By analyzing the first condition $\tilde{\zeta}-2-2\,j=2\,n$, we get the following expression of the energy eigenvalues $E_{n,l}$ :
\begin{equation}
E^2_{n,l}=k^2+M^2+2\,M\,\omega_c\,\left(n+1+\frac{l-\Phi}{\alpha}+\sqrt{\frac{(l-\Phi)^2}{\alpha^2}+M^2\,\Omega^2\,b^2_{2}+k^2_{c}-e^2\,\xi^2_{c}}\right).
\label{c8}
\end{equation}

The radial wave-function is given by
\begin{equation}
\psi_{n,l} (x)=x^{\sqrt{\frac{(l-\Phi)^2}{\alpha^2}+M^2\,\Omega^2\,b_2^{2}+k^2_{c}-e^2\,\xi^2_{c}}}\,e^{-\frac{x^2}{2}}\,H (x),
\label{c9}
\end{equation}
where $H (x)$ is a polynomial of degree $n$.

As done earlier, one can find here also the individual energy level and wave-function for each radial mode. For example, $n=1$, we have $\tilde{\zeta}-2-2\,j=2$ and $c_2=0$, hence from Eq. (\ref{c7})
\begin{eqnarray}
&&\tilde{\eta}\,c_1-2\,c_0=0\Rightarrow c_1=\frac{2}{\tilde{\eta}}\,c_0\Rightarrow \frac{\tilde{\eta}}{1+2\,j}=\frac{2}{\tilde{\eta}}\nonumber\\ \Rightarrow &&\omega^{1,l}_c=\left[\frac{a^2_{1,l}}{2\,M\,(1+2\,j)}\right]\nonumber\\ \Rightarrow &&B^{1,l}_0=\frac{1}{e}\,\left[\frac{a^2_{1,l}}{(1+2\,j)}\right]
\label{c10}
\end{eqnarray}
a constraint on the magnetic field $B^{1,l}_0$ so that first order polynomial solution is obtained.

For the radial mode $n=1$, one can evaluate the energy level $E_{1,l}$ from the following equation 
\begin{equation}
E^2_{1,l}=k^2+M^2+2\,M\,\omega_c^{1,l}\,\left(n+1+\frac{l-\Phi}{\alpha}+\sqrt{\frac{(l-\Phi)^2}{\alpha^2}+M^2\,\Omega^2\,b^2_{2}+k^2_{c}-e^2\,\xi^2_{c}}\right).
\label{energy}
\end{equation}
And The corresponding ground state wave-function is
\begin{equation}
\psi_{1,l}(x)=x^{\sqrt{\frac{(l-\Phi)^2}{\alpha^2}+M^2\,\Omega^2\,b_2^{2}+k^2_{c}-e^2\,\xi^2_{c}}}\,e^{-\frac{x^2}{2}}\,(c_1\,x+c_0),
\label{c11}
\end{equation}
where
\begin{equation}
c_1=\frac{1}{\sqrt{\left (\frac{1}{2}+\sqrt{\frac{(l-\Phi)^2}{\alpha^2}+M^2\,\Omega^2\,b_2^{2}+k^2_{c}-e^2\,\xi^2_{c}}\right)}}\,c_0.
\label{c12}
\end{equation}
Then, by substituting $\omega^{1,l}_c$ from Eq. (\ref{c10}) into the Eq. (\ref{energy}), it is possible to obtain the allowed value of the relativistic energy level $E_{1,l}$ for the radial mode $n=1$. We can see that the lowest energy state defined by the condition Eq. (\ref{c10}) plus the expression given in Eq. (\ref{c11}) is for the radial mode $n=1$, instead of $n=0$. This effect arise due to the presence of Coulomb-type potentials in the relativistic quantum system. Also the wedge parameter $\alpha$ are in the ranges $0 < \alpha <1$, thus, the degeneracy of the relativistic energy eigenvalue is broken and shifted the energy level of the bound states.

\section{Conclusions}

We have presented a detail study of generalized KG-oscillator interacts with an electromagnetic field in the presence of a scalar and vector potentials of Coulomb-types in cosmic string space-time. Through the minimal substitution, we have introduced an electromagnetic four-vector potential $A_{\mu}$ and a modification of the mass term $M \rightarrow M+ S (r)$ where, $S (r)$ is the scalar potential in the equation. We have solved generalized KG-oscillator equation under the effects of above potentials in the presence of a uniform external magnetic field $B_0$ including quantum flux $\Phi_B$ in a static cosmic string space-time. For the studies of generalized KG-oscillator, we have chosen different form of the arbitrary function $f (r)$, namely, Cornell-type, linear and Coulomb-type function. In {\it sub-section 2.1} we have chosen a Cornell-type function $f(r)$ in the equation and finally obtained the non-compact expression of the relativistic energy by the Eq. (\ref{22}) and the corresponding wave-functions Eq. (\ref{function}). By imposing the additional recurrence condition $c_{n+1}=0$ on the eigenvalue as done in \cite{AVV,JM}, one can obtain the ground state energy level $E_{1,l}$ from the Eq. (\ref{24}) and the ground state wave-function is given by Eqs. (\ref{25}--(\ref{26}). In {\it sub-section 2. 2}, we have chosen a linear function $f(r)$ and obtained the non-compact expression of the energy by the Eq. (\ref{b3}) and the corresponding wave-functions Eq. (\ref{b4}). By imposing the recurrence condition $c_{n+1}=0$, we have obtained the Eq. (\ref{b5}) from which one can find the ground state energy $E_{1,l}$ defined for the radial mode $n=1$, and the corresponding wave-function Eqs. (\ref{b6}--(\ref{b8}). In {\it sub-section 2. 3}, we have chosen a Coulomb-type function $f(r)$ and obtained the non-compact expression of the energy by the Eq. (\ref{c8}) and the corresponding wave-functions Eq. (\ref{c9}). Here also, using the recurrence condition we have obtained Eq. (\ref{energy}) for the radial mode $n=1$ from which one can find the ground state energy $E_{1,l}$, and the corresponding wave-function is given by Eqs. (\ref{c11})--(\ref{c12}). In {\it sub-section 2.1 --2.3}, we have seen that the presence of Coulomb-types scalar and vector potentials allow the formation of bound states solution of the considered relativistic system and hence, the lowest energy state is defined by the radial mode $n=1$, instead of $n=0$. Also in gravitation and cosmology, the values of the wedge parameter $\alpha$ are in the ranges $0 < \alpha <1$, and thus, the degeneracy of the energy eigenvalue is broken and shifted the relativistic energy level. An important observation when we tries to analyze $c_{n+1}=0$ for the radial mode $n=1$ is noted in all cases, where certain parameter is constraint, for example, $\omega_{1,l}$ that appears in {\it sub-section 2.1} by the Eq. (\ref{23}) depend on the parameter $a_{1,l}$ which is already depend on energy $E_{1,l}$ by the Eq. (\ref{11}) because of the presence of Coulomb-type electromagnetic potential considered in this quantum system. Another interesting observation that we made in this work is the quantum effect which arises due to the dependence of the magnetic field $B^{n,l}_0$ on the quantum number $\{n, l\}$ of the relativistic quantum system. 

 In {\it sub-section 2.1 --2.3}, we have observed that the individual energy level for each radial mode depends on the quantum flux and thus is a periodic function of the geometric quantum phase $\Phi_B$ because the angular quantum number $l$ is shifted, $l \rightarrow l_0=\frac{1}{\alpha}\,\left(l-\frac{\Phi_B}{(2\pi/e)} \right)$, an effective angular momentum quantum number. Hence, we have that, $E_{n,l} (\Phi_B+\Phi_0)=E_{n, l \mp \tau} (\Phi_B)$ where,  $\Phi_0=\pm\,\frac{2\,\pi}{e}\,\tau$ with $\tau=0,1,2,...$. This dependence of the relativistic energy level on the geometric quantum phase gives us the relativistic analogue of the Aharonov-Bohm effect for bound states. So, when the particle circles around the string, the total wave-function $\Psi$ changes to a new function given by $\Psi \rightarrow \Psi'=e^{2\,\pi\,i\,l_0}\,\Psi=e^{\frac{2\,\pi\,i}{\alpha}\,\left (l-\frac{e\,\Phi_B}{2\,\pi}\right)}\,\Psi$.

\section*{Acknowledgement}

Author sincerely acknowledged the anonymous kind referee for his/her valuable comments and suggestions.

\section*{Conflict of Interest}

Author declares that there is no conflict of interest regarding publication this paper.

\end{document}